# Non-Volatile Resistive Switching of Polymer Residues in 2D Material Memristors

*Dennis Braun[1], Mohit D. Ganeriwala[2], Lukas Völkel[1], Ke Ran[3,4], Sebastian Lukas[1], Enrique G. Marín[2], Oliver Hartwig[5], Maximilian Prechtl[5], Thorsten Wahlbrink[6], Joachim Mayer[3,4], Georg S. Duesberg[5], Andrés Godoy[2], Alwin Daus[1,\*] and Max C. Lemme[1,6,\*]*

[1]Chair of Electronic Devices, RWTH Aachen University, Otto-Blumenthal-Str. 2, 52074 Aachen, Germany.

[2]Department of Electronics and Computer Science, Universidad de Granada, Avenida de la Fuente Nueva S/N 18071, Granada, Spain.

[3]Central Facility for Electron Microscopy (GFE), RWTH Aachen University, Ahornstr. 55, 52074 Aachen, Germany.

[4]Ernst Ruska-Centre for Microscopy and Spectroscopy with Electrons (ER-C 2), Forschungszentrum Jülich, 52425 Jülich, Germany.

[5]Institute of Physics, Faculty of Electrical Engineering and Information Technology (EIT 2) and Center for Integrated Sensor Systems, University of the Bundeswehr Munich, 85577 Neubiberg, Germany.

[6]AMO GmbH, Advanced Microelectronic Center Aachen, Otto-Blumenthal-Str. 25, 52074 Aachen, Germany.

[\*]Corresponding authors' emails: max.lemme@eld.rwth-aachen.de; alwin.daus@eld.rwth-aachen.de

**Abstract**

Two-dimensional (2D) materials are popular candidates for emerging nanoscale devices, including memristors. Resistive switching (RS) in such 2D material memristors has been attributed to the formation and dissolution of conductive filaments created by the diffusion of metal ions between the electrodes. However, the area-scalable fabrication of patterned devices involves polymers that are difficult to remove from the 2D material interfaces without damage. Remaining polymer residues are often overlooked when interpreting the RS characteristics of 2D material memristors. Here, we demonstrate that the parasitic residues themselves can be the origin of RS. We emphasize the necessity to fabricate appropriate reference structures and employ atomic-scale material characterization techniques to properly evaluate the potential of 2D materials as the switching layer in vertical memristors. Our polymer-residue-based memristors exhibit RS typical for a filamentary mechanism with metal ion migration, and their performance parameters are strikingly similar to commonly reported 2D material memristors. This reveals that the exclusive consideration of electrical data without a thorough verification of material interfaces can easily lead to misinterpretations about the potential of 2D materials for memristor applications.

**Introduction**

Resistive switching (RS) is observed in many materials like metal-oxides[1], polymers[2], phase-change materials[3] and 2D materials[4]. Electronic devices made from RS materials can be switched between two or more resistive states in a volatile or non-volatile way and are often termed "memristors" or described as having "memristive behavior"[5–9]. Memristors are currently being investigated for their potential to support neuromorphic computing, e.g. for enabling computing-in-memory[10], cross-bar arrays[11], or to act as electronic synapses[12]. Their arrangement in crossbar arrays is preferred for application-relevant implementations, and many studies employ single crosspoint devices to study device-level metrics of single cells[5]. They typically consist of two conducting electrodes separated by the RS material (e.g. schematics in Fig. 1a).

2D layered materials have recently received attention as resistive switching materials because they potentially enable fast and low-power switching, large switching windows, forming-free characteristics, utilization on arbitrary substrates, and ultimate thinness. Two-terminal, vertical devices have been demonstrated from 2D hexagonal boron nitride[4,11,13–16], transition metal dichalcogenides[15,17–21], noble metal dichalcogenides[15,22,23] and 2D heterostructures[15,24,25]. Several of these pioneering studies[4,13,17] completely avoided polymers during device fabrication by using direct growth of 2D materials, common back-electrodes, and shadow masks for metal deposition. These fabrication techniques, however, are usually not scalable, i.e., they cannot be adopted for future mass manufacturing. Furthermore, the growth temperature of most 2D materials is, with a few recent exceptions[26,27], too high to grow them directly on CMOS circuitry, which is the envisioned strategy for integrating 2D materials into semiconductor processing lines[28–30]. Conventional scalable fabrication processes for crosspoint devices from 2D materials typically involve several steps that use photoresists and other polymers, including patterning the back- and top-electrodes with lift-off or etching techniques, transferring the 2D material from the growth substrate onto the back-electrode, and patterning the 2D material to isolate the individual cells and prevent lateral leakage paths. This presents a potential challenge for vertical devices because polymer residues are a commonly observed issue[5,31–38] for 2D materials and may be incorporated into the active device stack. In addition, dry etching, hard baking, and metal coating are known to cross-link photoresists either thermally or optically by deep-ultraviolet radiation, making them harder to remove[39]. Finally, polymer residues cannot be cleaned from 2D materials as aggressively as from silicon.

Consequently, understanding the role of polymer residues inside 2D memristors is crucial for understanding 2D material switching characteristics. Several studies have shown cross-sectional transmission electron microscopy (TEM) analyses of their vertical memristors to assess the cleanliness of the interfaces and exclude parasitic layers inside the active structures[4,11,14,16,17,24,25]. However, there are also studies that use polymers for the fabrication and do not provide TEM images of their final device stacks[13,18,20,40–43] and at least two studies that appear to have parasitic, amorphous layers inside the device with unknown roles in the RS[22,23].

In this study, we demonstrate reproducible RS in vertical memristors with metal electrodes using thin polymer residues as the active material, which were created during the patterning of the 2D material platinum diselenide (PtSe$_2$). We find that devices containing only polymer, but excluding the 2D material in the fabrication process, switch similarly to PtSe$_2$ devices that include polymer residues from the fabrication. Such polymer-residue memristors operate with comparable voltages and display similar resistance ranges to 2D material memristors in the literature. Area-dependent measurements and modeling of our experimental data indicate a filamentary switching mechanism, analogous to conductive bridge random access memories (CBRAM), a mechanism also proposed for many 2D material memristors[4,15,16,44–46]. These results highlight that RS in polymer residues and 2D materials can easily be confused.

**Device Fabrication**

We fabricated vertical crosspoint and crossbar array memristors based on ~3.7 nm and ~12 nm thick $PtSe_2$ with palladium (Pd) back-electrodes (BE) and nickel (Ni) top-electrodes (TE). The samples are named $PtSe_2$-A and $PtSe_2$-B, respectively, and Fig. 1a depicts their schematic device structures and microscope images. A third sample was fabricated using the same processing steps but without the 2D material (polymer sample, Fig. 1a). Atomic force microscopy (AFM) of the as-grown $PtSe_2$, Raman spectra confirming the presence of layered $PtSe_2$ through the spectral fingerprint, AFM analyses of the BE, and additional micrographs of the devices are provided in the supplementary information (SI) Fig. S1. Details on the fabrication process are given in the Methods section. Cross-sectional transmission electron microscopy (TEM) imaging revealed an amorphous ~5-7 nm thick layer between $PtSe_2$ and the Ni TE in both $PtSe_2$ samples (Figs. 1b and c). Energy-dispersive X-ray spectroscopy (EDX) analyses of the elements show the onset of a carbon signal in the gap between the Ni-TE and the $PtSe_2$ peaks (Figs. 1d and e), strongly suggesting the amorphous layer to be polymer residues from fabrication (see Methods for details). The lower intensity of the carbon peaks, when compared to the other elements, is related to the difficulty of detecting low atomic number elements like carbon due to their low X-ray energies[47]. Additionally, the overlap of the Pt, Se, C and Pd peaks in Fig. 1e can be explained by the redeposition of material during the focused ion beam milling of the TEM lamellas[48]. The expected layered structure of PtSe$_2$ (Raman spectra in SI Fig. S1d) is not visible in the TEM images (Fig. 1b,c). We attribute this to the nanocrystallinity of the material with crystallites tilted in various directions, rendering them invisible under certain angles in the TEM[49]. AFM imaging of the active area of the polymer sample before TE deposition shows a polymer thickness of ~22 nm (Fig. 1f,g). Additional TEM images and EDX maps of samples PtSe$_2$-A and -B, and AFM scans of the active regions of the PtSe$_2$ samples before the TE deposition are provided in the SI Figs. S2, S3 and S4.

**Resistive switching in polymer residues**

All samples described in Fig. 1 were subjected to direct-current (dc) current-voltage (*I-V*) cycles, where the Ni TE voltage was swept from $V_{TE}$ = 0 V to the positive SET voltage and back to 0 V, then from 0 V to the negative RESET voltage and finally back to 0 V (Fig. 2a-c). The devices showed bipolar non-volatile RS within a voltage range of -1.3 V to 1.5 V and switched between distinct high resistive states (HRS) and low resistive states (LRS). Details on the *I-V* measurements can be found in the Methods section, while time-dependent *I-V* data of a cycle is plotted in SI Fig. S5a. The presence of parasitic metal oxides was excluded with reference *I-V* measurements on devices with a direct contact between the BE and the TE (SI, Fig. S5b). Samples PtSe$_2$-A and PtSe$_2$-B exhibit similar *I-V* curves (Fig. 2a,b), despite the different PtSe$_2$ film thicknesses and the different morphology of the films visible in the TEM images (PtSe$_2$-A shows island-like grains while PtSe$_2$-B shows a more closely packed film, see Fig. 1b,c). However, the samples have residual polymer layers of similar thicknesses, which indicates that the polymer dominates the switching and not the PtSe$_2$. This is consistent with the *I-V* measurements on the polymer sample that shows RS without PtSe$_2$ in the device stack (Fig. 2c). Hence, we conclude that AZ5214E photoresist (MicroChemicals) is the switching medium in all three samples. This resist is the top layer of the multi-layer lithography process that was used for patterning the PtSe$_2$ and for the active region of the polymer sample (see Methods section) and thus was exposed to the etching plasma and the ultraviolet radiation that both led to cross-linking and hardening[39]. We compare the distribution of the HRS and LRS resistances and the SET and RESET voltages of four devices each for PtSe$_2$-A (blue) and -B (green) and five devices for the polymer sample (red) in Fig. 2d and e. The similarities in samples PtSe$_2$-A and -B are consistent over all devices and cycles for both switching voltages and LRS and HRS resistances. Moreover, the LRS resistances and both SET and RESET voltages are very similar for all three samples, indicating a common switching mechanism, which confirms the polymer as the switching medium. We attribute the higher HRS resistances in the polymer sample to the thicker polymer layer in the device stack. The higher HRS variability can be explained by the larger number of wrinkles on the surface of the active material in the polymer sample compared to both PtSe$_2$ samples (see AFM scans before the TE deposition in Fig. 1f,g and SI Fig. S4). Kim et al.[50] recently reported a PtSe$_2$ insertion layer leading to lower variations in the threshold voltages and better endurance of volatile hafnium dioxide (HfO$_2$) threshold memristors. They link this result to thinner and weaker conductive filaments formed on the PtSe$_2$ layer compared to an inert metal electrode due to the low work function, superior inertness, and layered structure of PtSe$_2$. We propose a similar mechanism here in accordance with the proposed filamentary switching mechanism. The filaments through the polymer in samples PtSe$_2$-A and -B are weaker and thinner because they contact the PtSe$_2$ layer instead of the Pd BE in the polymer sample. Consequently, they dissolve more easily, leading to less variation in the HRS.

The filaments in the polymer sample without PtSe$_2$ are thicker and more stable and hence harder to dissolve fully, which leads to the HRS spreading over up to four orders of magnitude.

We have summarized our device statistics in Fig. 2f,g and compared them to state-of-the-art 2D material memristors. The boxplots for PtSe$_2$-A, -B and the polymer sample summarize all devices plotted in Fig. 2d,e (four devices with a total of 95 *I-V* cycles, four devices with a total of 86 cycles, and five devices with a total of 222 cycles, respectively). Our results clearly overlap with the results on 2D material devices in the literature. We compare our results to studies that provide TEM analysis of their devices, which show clean interfaces and where transfer-free methods are used[4,17], to studies that did not provide TEM analysis[13,20], and also to studies where TEM imaging shows amorphous layers in the device stack[22,23]. At least three of the four parameters, HRS, LRS, SET, and RESET voltage, overlap with our results for all mentioned publications. This shows that the RS characteristics of our polymer layer are very similar to the RS characteristics of 2D material memristors. Additionally, these similarities demonstrate that the *I-V* characteristics alone cannot provide proof of the origin of the RS behavior as long as a residual polymer layer cannot be avoided with certainty.

We analyzed the electrical characteristics of our polymer sample in more detail. Electrode-area-dependent measurements show an area-dependent HRS and an area-independent LRS (Fig. 3a), indicating a filamentary resistive switching mechanism. We, therefore, assume that few nano-scale metallic filaments carry the current in the LRS, while the filaments partially dissolve so that the entire active area contributes to current conduction in the HRS[51,52]. We further fitted the current conduction mechanism of the median *I-V* curves of several devices for the HRS and the LRS to a transport model. In the HRS, the current through the polymer as a function of the applied voltage is well described by the space charge limited conduction (SCLC) mechanism[53]

$$I = 4\pi^2 \frac{k_B T}{q} \mu \epsilon A \frac{V}{L^3} + \frac{9}{8} \mu \epsilon A \frac{V^2}{L^3} \quad (1)$$

Here, *I* is the current, $k_B T$ *is* the thermal energy, *q* is the elementary charge, *μ is* the charge carrier mobility, $\epsilon$ is the permittivity of the conductive medium, and *A* is the cross-sectional area. Equation (1) describes a linear regime (*I* ∝ *V*) for lower voltages and the Mott-Gurney square law (*I* ∝ $V^2$) at higher voltages. These trends are mirrored in the slope of the median *I-V* curve of a device in the HRS in Fig. 3b. The current in the LRS exhibits a linear *I-V* dependence (Fig. 3c) in agreement with Ohmic conduction via metallic filaments bridging the electrodes through the polymer. In Fig. 3d, we fitted both Ohmic conduction and SCLC via Equation (1) to the full *I-V* curve, choosing appropriate pre-factors (see SI Fig. S6 for details), excluding the compliance and transition regions. There is good agreement between the model and the experimental data. We propose that the TE is oxidized at a positive voltage, releasing $Ni^+$ ions which drift towards the BE(Pd), where they are reduced and deposited. We assume that the initial nucleation and growth of the Ni filaments happens at the most conductive locations within the device[54], e.g. where the polymer is the thinnest. This results in the simultaneous growth of multiple Ni filaments at the BE (Fig. 3e). In the HRS, none of the filaments bridge the electrodes. Hence carrier injection happens via multiple growing filaments, acting as a virtual BE, leading to an area-dependent current. Eventually, some Ni filaments will bridge the gap between the electrodes (Fig. 3f), resulting in several low-resistance metallic paths through the polymer, switching the conduction mechanism from SCLC to Ohmic. Due to the unequal filament growth speed, only a few filaments will bridge the electrodes. These filaments then carry all the current, making the LRS area independent. SI Fig. S6 shows the median HRS and LRS curves of four additional devices, which confirm the observed trends. In the HRS, most of the applied voltage drops across the gap between the tip of the filaments and the TE since the Ni filaments act as a virtual BE. The $L$ in eq. (1) is, therefore, equivalent to the average gap size from multiple filaments. During the SET transition, while SCLC still governs the current, the gap size reduces due to filament growth. Consequently, $L$ in eq. (1) becomes voltage-dependent, giving rise to higher than $V^2$ dependence in the current. This can explain the sharp rise in the current during the SET-transition visible in all *I-V* curves (Fig. 2a-c) that we emphasize in Fig. 3g with a median *I-V* curve for four devices in the polymer sample. This filamentary switching mechanism (CBRAM) is similar to the proposed mechanisms for many 2D material devices[4,15,44–46], which explains the similarity between the RS in our polymer residues and 2D materials in the literature. Moreover, the thicknesses of the polymer films in PtSe$_2$-A, -B and the polymer sample are akin to those of multilayer 2D materials, which are frequently favored over monolayers for memristive devices[11]. The diffusion of metal ions from the electrode into an insulating, nanometer-scale film and the subsequent formation of metallic filaments that bridge the insulating film leads to a characteristic RS behavior that can be demonstrated in vastly different material systems. Although RS in polymers is an active research area itself[2,55–60], most of the polymers reported are not as thin as in our samples since they are deposited by spin-coating without further processing.

**Conclusions**

We demonstrated reproducible RS in vertical memristors based on parasitic polymer residues introduced during the processing of 2D materials. We show that these polymer residues dominate the switching behavior over the 2D material. The voltage and resistance ranges of the RS in our polymer devices are similar to those in state-of-the-art 2D material memristors. Our analysis and modeling indicate a filamentary RS mechanism, which has also been associated with 2D material memristors in the literature. We conclude that thorough electrical and material analyses, including appropriate reference structures, are necessary when investigating RS in 2D material memristors to distinguish it from RS in parasitic residual polymer layers. This is particularly important because the fabrication of 2D material-based devices with conventional CMOS fabrication techniques often leads to polymer residues, especially in vertical device configurations.

**Methods**

**Device Fabrication**

2x2 cm$^2$ highly p-doped silicon (Si) chips with 300 nm silicon dioxide (SiO$_2$) grown by thermal oxidation on top as insulation were used as substrates. The back-electrode (BE) was patterned via a negative optical lithography process using a double layer of lift-off resist (LOR 3A, micro resist technology) and AZ5214E photoresist (MicroChemicals). Then, 10 nm of titanium and 40 nm of palladium were deposited by electron-beam evaporation, followed by lift-off in dimethyl sulfoxide (DMSO). All samples presented in this work were fabricated with the same BE process.

Platinum diselenide (PtSe$_2$) was grown by thermally assisted conversion of thin, sputtered Pt films on Si/SiO$_2$ chips at 450°C in a furnace using Se vapor[26,49,61]. Crosspoint and crossbar array devices with PtSe$_2$ were fabricated by transferring the respective PtSe$_2$ film (A or B) from the growth substrate to the patterned BE via a poly(methyl methacrylate) (PMMA) assisted wet transfer process. The material was delaminated from the growth substrate with a potassium hydroxide solution. To pattern the PtSe$_2$ in the active device region a positive lithography step using a double layer of LOR 3A and AZ5214E photoresist on top of the PMMA[62] from the transfer was performed. Hence, after development the PtSe$_2$ was still covered by PMMA from transfer in the region supposed to be etched. Low power reactive ion etching (ICPRIE) with oxygen plasma was used to etch the PMMA, then PtSe$_2$ was etched in a chlorine, oxygen and argon plasma. The LOR 3A and AZ5214E resists were stripped in DMSO at room temperature[62] for 10 minutes, then the PMMA was stripped in acetone at 65 °C for 20 minutes. The polymer-only memristors were fabricated by first spin coating PMMA onto the BE (instead of the transfer of PtSe$_2$), then the lithography, etching and resist stripping steps were the same as for the PtSe$_2$ samples.

The top-electrode (TE) was patterned by a negative lithography process using AZ5214E with subsequent dc sputtering of 55 nm of nickel and 15 nm of aluminum and lift-off in acetone. All samples presented in this study were fabricated using this process.

**Material and Device Characterization**

Cross-sectional transmission electron microscope (TEM) and energy dispersive X-ray spectroscopy (EDX) investigations were conducted with a JEOL-JEM200 TEM at 200 kV operation voltage. Thin lamellas for TEM and EDX analysis were cut from the respective crosspoint or crossbar devices with a Strata400 focused ion beam system using a gallium ion beam. Atomic force microscopy (AFM) measurements were performed using a Dimension Icon AFM by Bruker in tapping mode. Raman spectroscopy was done with a WiTec alpha300R Raman imaging microscope using an excitation laser wavelength of 532 nm with 1 mW laser power.

**Electrical characterization**

Electrical measurements were conducted in a cryogenic probe station (LakeShore) connected to a semiconductor parameter analyzer (SPA) 4200A-SCS by Tektronix at room temperature in ambient conditions. Two source measure unit (SMU) cards (Keithley 4200-SMU), each connected to a pre-amplifier (Keithley 4200-PA), were used. Repeated voltage cycles were applied to the TE, while the BE was grounded, according to the sequence indicated by the arrows and numbers in the graphs in Figs. 2a-c. Firstly, a voltage sweep from 0 V to the positive SET voltage was applied during which the device switched from the high-resistive state (HRS) to the low-resistive state (LRS). The internal current compliance of the semiconductor parameter analyzer was used to limit the current during the SET transition to avoid permanent breakdown. The current compliances used in the exemplary *I-V* curves in Figs. 2a-c were 300 µA for PtSe$_2$-A and 100 µA for PtSe$_2$-B and the polymer sample, respectively. We found no influence of the current compliance on the resistance levels or switching within this range. The voltage was automatically decreased as soon as the current reached compliance during the SET transition, which leads to a part of the curve where the voltage decreases while the current increases (*I-V-t* plot in the SI Fig. S5a). Secondly, the voltage was swept back from the SET voltage to 0 V while the device remained in the LRS. The current remained in compliance during most of the backward sweep due to the low resistance in the LRS. Thirdly, a voltage sweep from 0 V to the negative RESET voltage was applied, during which the device switched back from LRS to HRS. The current compliance was deactivated for the RESET sweep to reach the currents required to induce the transition. Lastly, the voltage was swept back from the RESET voltage to 0 V while the device remained in the HRS. The high and low resistive state resistances were measured with a constant dc voltage of 10 mV before and after the respective SET or RESET sweeps.

**Acknowledgements**

Financial support from the German Federal Ministry of Education and Research, BMBF, within the projects NEUROTEC (Nos. 16ES1134 and 16ES1133K), NEUROTEC 2 (Nos. 16ME0399, 16ME0398K, and 16ME0400), NeuroSys (No. 03ZU1106AA and 03ZU1106BA) and NobleNEMS (No. 16ES1121K) is gratefully acknowledged.

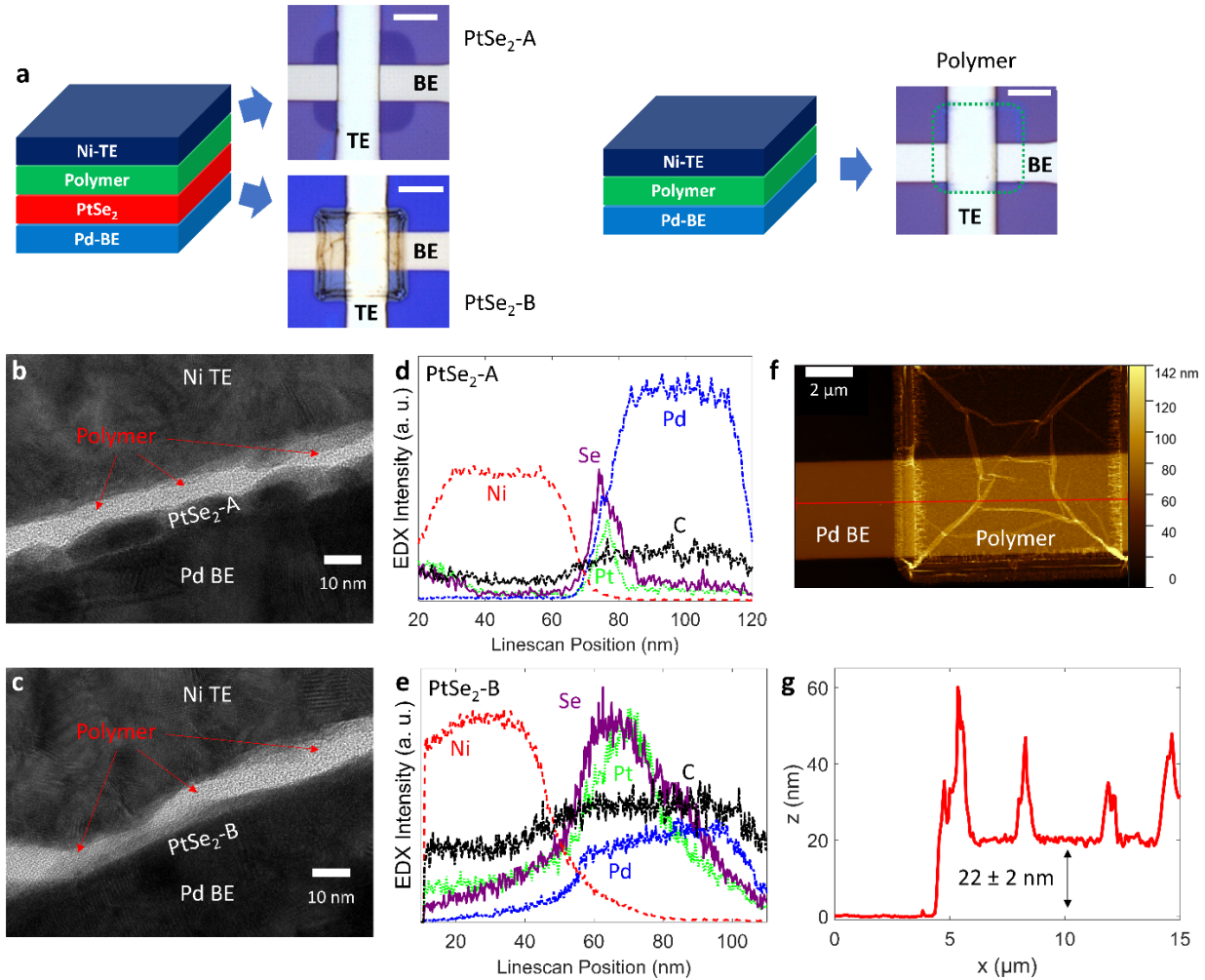

Figure 1: Memristor device stacks and structural characterization. a) Schematic cross-sections and top-view optical images of the three samples fabricated in this study. The patterned active material of the polymer sample is marked with a green, dashed line. The scale bar in each optical image represents 5 µm. b) Cross-sectional TEM image of PtSe$_2$-A, showing the top- and bottom-electrodes, PtSe$_2$-A (average thickness 4.8 ± 1.2 nm) and the amorphous residual polymer from etching (marked by the red arrows, average thickness 5.4 ± 1.3 nm). c) TEM image of PtSe$_2$-B (average thickness 11.3 ± 1.7 nm) showing the intentional layers and the amorphous polymer residue (marked by the red arrows, average thickness 7.5 ± 1.7 nm). d, e) EDX intensities of line

scans of PtSe$_2$-A and -B showing the elements contained in each layer. The polymer appears as an increased carbon signal in the device stack. f) AFM scan of the active area of a polymer-only device before the TE deposition. The Pd BE and the polymer pad are visible. A polymer thickness of 22 ± 2 nm is deducted from the height profile marked in red, see subfigure g). g) AFM height profile along the red line in subfigure d). The thickness values for PtSe$_2$-A and -B and the respective polymer layers were derived from the TEM images depicted here and in SI Fig. S2 by averaging at least 15 individual measurements per image using the software ImageJ.

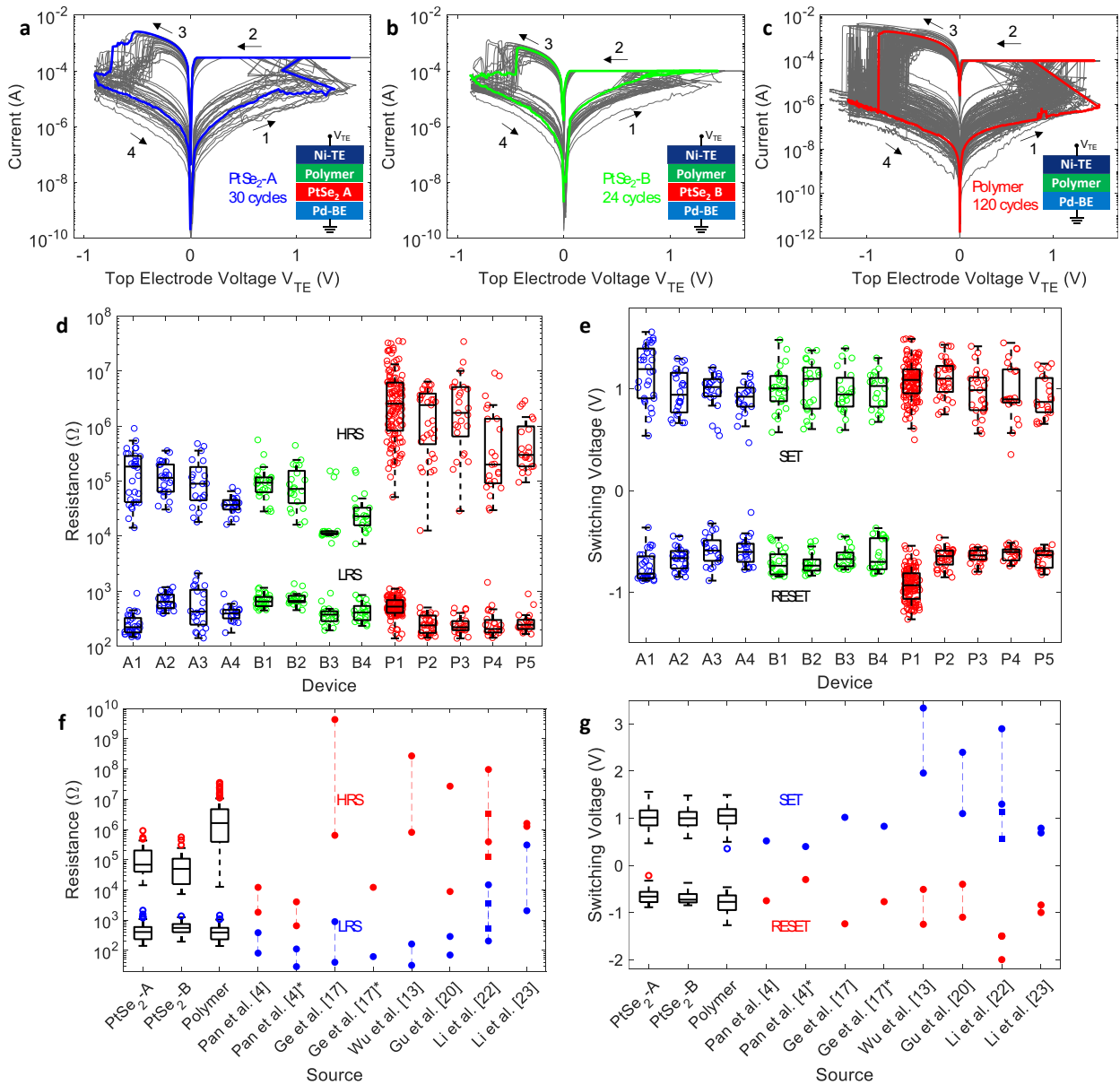

Figure 2: Electrical dc *I-V* characterization of PtSe$_2$-A, -B and polymer samples and comparison to 2D memristors associated with a filamentary switching mechanism in the literature[4,13,17,20,22,23] a) 30 switching cycles of a memristor in sample A. The arrows indicate the voltage sweep sequence. The inset schematic shows the device layers and the biasing scheme. The first sweep is colored blue. b) 24 switching cycles of a memristor in sample B. The first sweep is colored green. c) 120 switching cycles of a memristor in the polymer sample. The first sweep is colored red. d) Boxplots

for the statistical comparison of HRS and LRS resistances of four devices each for $PtSe_2$-A and -B (devices A1 to A4 and B1 to B4 in blue and green, respectively) and five devices for the polymer sample (devices P1 to P5 in red). e) Boxplots for the statistical comparison of SET and RESET voltages. f) Statistical comparison of the HRS and LRS resistances (colored in red and blue, respectively) of our samples with state-of-the-art 2D material based memristors in the literature that are associated with a filamentary switching mechanism. For the literature data the vertically connected data points were taken from a provided distribution plot of the respective reference, representing the maximum and the minimum values observed. Single data points were extracted from *I-V* curves or written data when the reference did not include a distribution. Data marked with an asterisk (*) means that the corresponding device was fabricated without the use of photolithography and transfer, thus ruling out polymer contamination. g) Comparison of the SET and RESET voltages (colored in blue and red, respectively) of our samples with 2D memristors published in literature.

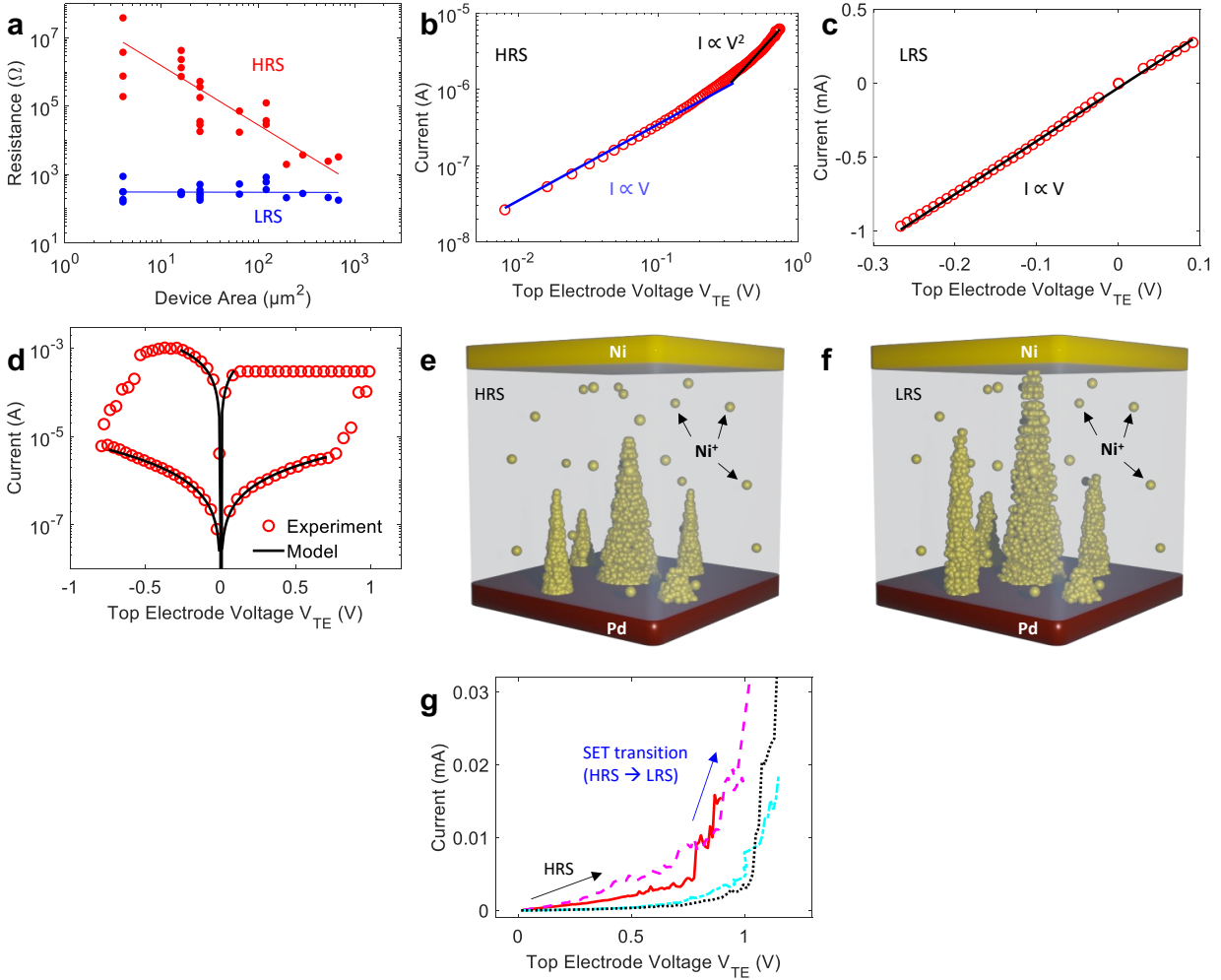

Figure 3: Modeling of the resistive switching mechanism in the polymer sample. a) Double logarithmic plot of HRS and LRS of devices with a variable active area (676 µm$^2$ down to 4 µm$^2$) and linear fits of the log-log data. Each data point is the average of at least 5 full dc *I-V* sweep cycles in a single device. The LRS resistance stays constant with an increasing active area, while the HRS resistance increases with a decreasing active area, indicating a filamentary switching mechanism. b) Double logarithmic plot showing a median *I-V* curve of a device in the HRS with a linear and a quadratic fit in correspondence to the two *I-V* trends of SCLC (equation 1). c) Median *I-V* characteristic of a device in the LRS with a linear fit, confirming Ohmic conduction. d) Full median *I-V* sweep with the modeling fits for both states. The HRS is modeled by SCLC and the LRS

by Ohmic conduction, showing good agreement with the data. The current compliance and the transition between the states are excluded from this analysis. e) Schematic representation of the device in the HRS. f) Schematic representation of the device in the LRS. g) Median *I-V* plots of four devices showing the sharp current increase during the SET process (transition region), that we explain by SCLC with reducing filament gap size.

# Supplementary Information

# Non-Volatile Resistive Switching of Polymer Residues in 2D Material Memristors

*Dennis Braun[1], Mohit D. Ganeriwala[2], Lukas Völkel[1], Ke Ran[3,4], Sebastian Lukas[1], Enrique G. Marín[2], Oliver Hartwig[5], Maximilian Prechtl[5], Thorsten Wahlbrink[6], Joachim Mayer[3,4], Georg S. Duesberg[5], Andrés Godoy[2], Alwin Daus[1,\*] and Max C. Lemme[1,6,\*]*

[1]Chair of Electronic Devices, RWTH Aachen University, Otto-Blumenthal-Str. 2, 52074 Aachen, Germany.

[2]Department of Electronics and Computer Science, Universidad de Granada, Avenida de la Fuente Nueva S/N 18071, Granada, Spain.

[3]Central Facility for Electron Microscopy (GFE), RWTH Aachen University, Ahornstr. 55, 52074 Aachen, Germany.

[4]Ernst Ruska-Centre for Microscopy and Spectroscopy with Electrons (ER-C 2), Forschungszentrum Jülich, 52425 Jülich, Germany.

[5]Institute of Physics, Faculty of Electrical Engineering and Information Technology (EIT 2) and Center for Integrated Sensor Systems, University of the Bundeswehr Munich, 85577 Neubiberg, Germany.

[6]AMO GmbH, Advanced Microelectronic Center Aachen, Otto-Blumenthal-Str. 25, 52074 Aachen, Germany.

[\*]Corresponding authors' emails: max.lemme@eld.rwth-aachen.de; alwin.daus@eld.rwth-aachen.de

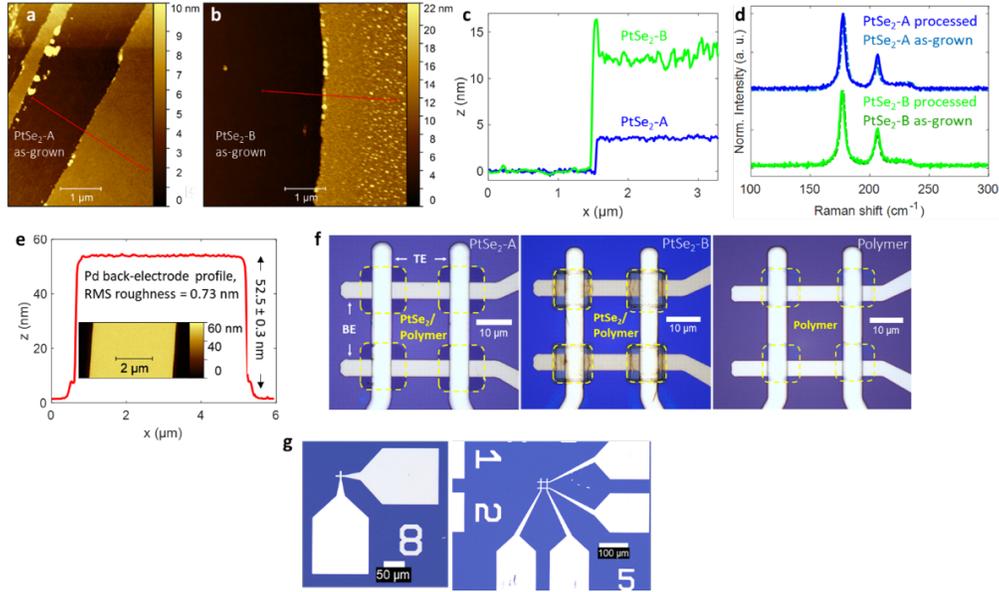

Figure S1: Additional structural characterization and microscope images. a) and b) show AFM images of the as-grown, polycrystalline $PtSe_2$-A and B, respectively. c) Height profiles along the red lines in a) and b) showing a $PtSe_2$ thickness of 3.7 ± 0.1 nm (A) and 12 ± 0.1 nm (B) on the growth substrate. The as-grown thicknesses correspond well with the cross-sectional TEM-determined thicknesses (Fig. 1b,c in the main text and supplementary Fig. S2). d) Raman spectra of both $PtSe_2$ samples, recorded as-grown and after transfer and processing, respectively, showing the characteristic peaks[1] of the material. There is no significant change due to processing. $PtSe_2$ was grown by thermally assisted conversion (TAC) of thin, sputtered Pt films on $Si/SiO_2$ at 450 °C in a furnace using Se vapor[1,2]. e) AFM scan and thickness profile of a typical Pd back-electrode showing a uniform rectangular shape, a thickness of 52.5 ± 0.3 nm, and a smooth surface with a root mean square roughness of about 0.73 nm. f) Zoomed-in optical images of 2x2 crossbar arrays of $PtSe_2$-A, -B and the polymer sample that have been used for the electrical characterization together with the crosspoint devices depicted in Fig. 1a. The patterned active material is marked with a yellow, dashed line g) Optical images of a full crosspoint and a 2x2 crossbar array device showing the contact pads for the probe station and the metal lines connecting them with the active region.

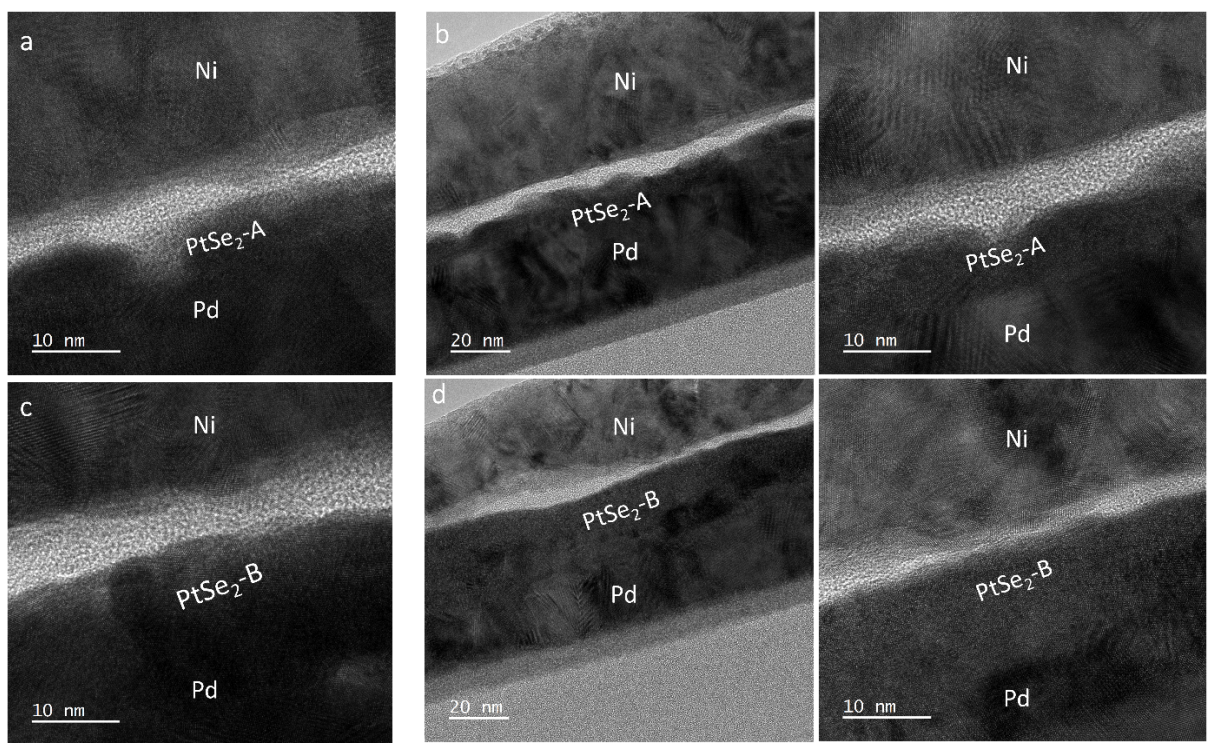

Figure S1: Additional TEM images. a) Magnified area of Fig. 1b in the main text. b) TEM image of a second PtSe$_2$-A device (zoomed in on the right). c) Magnified area of Fig. 1c in the main text. b) TEM image of a second PtSe$_2$-B device (zoomed in on the right). The images show Ni top- and Pd back-electrodes, the PtSe$_2$, which shows island-like grains in the case of sample A and a more closed, uniform layer for sample B, and the amorphous polymer layer for all samples in between the PtSe$_2$ and the top-electrode. The active material thicknesses from the TEM analysis are 4.8 ± 1.2 nm (PtSe$_2$-A) and 5.4 ± 1.3 nm (polymer on PtSe$_2$-A) and 11.3 ± 1.7 nm (PtSe$_2$-B) and 7.5 ± 1.7 nm (polymer on PtSe$_2$-B), respectively. The thickness values were derived from the TEM images by averaging at least 15 individual measurements per image using the software ImageJ.

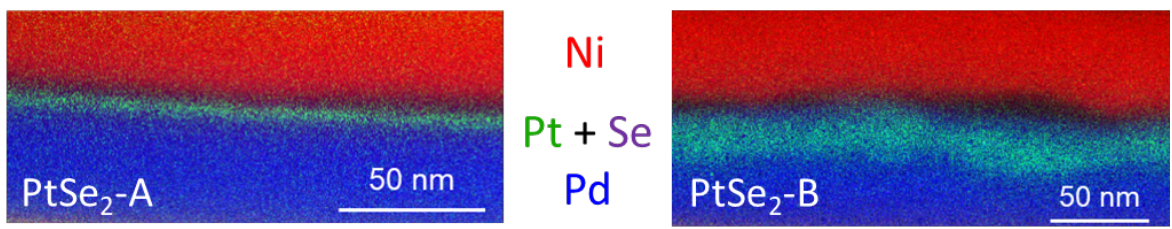

Figure S3: EDX mapping of PtSe$_2$-A and -B. The intensity map shows the intentional layers Ni, Pt, Se and Pd in the devices. The gap between Pt + Se and Ni indicates the position of the residual polymer layer.

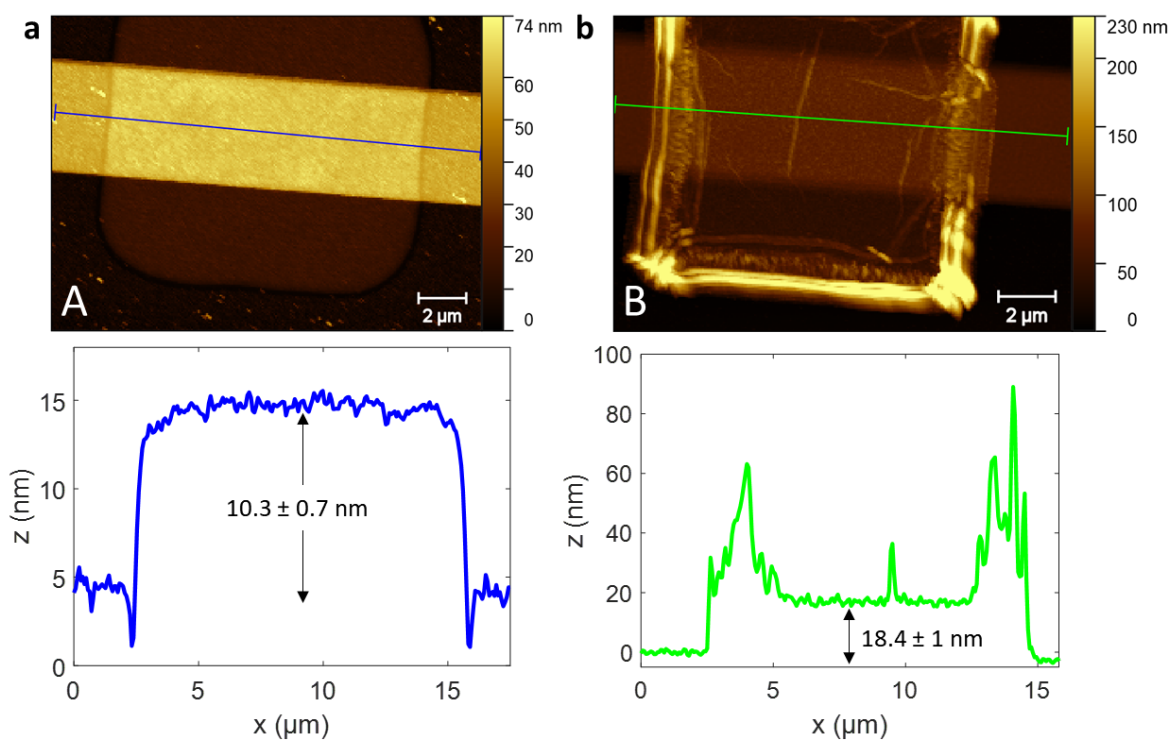

Figure S4: AFM scans and corresponding thickness profiles of the active device areas of PtSe$_2$-A (a) and PtSe$_2$-B (b) before the top-electrode was deposited. The thickness of the active material measured from the back-electrode is 10.3 ± 0.7 nm for PtSe$_2$-A and 18.4 ± 1 nm for PtSe$_2$-B, respectively, which corresponds well with the added PtSe$_2$ and polymer thickness from the TEM analysis (10.2 ± 1.8 nm for PtSe$_2$-A, 18.9 ± 2.5 nm for PtSe$_2$-B, see Fig. 1b,c in the main text and supplementary Fig. S2).

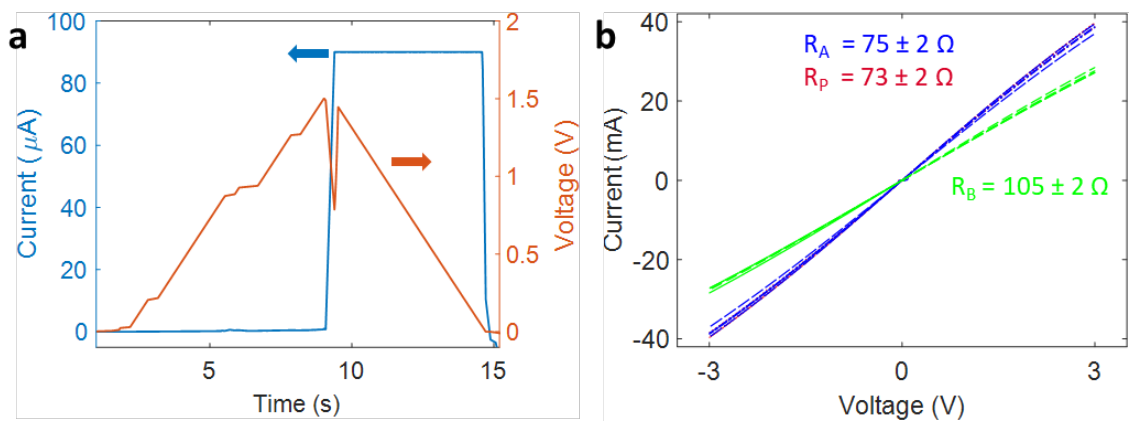

Figure S5: Additional electrical characterization data. a) Current and voltage vs. time for a typical dc *I-V* sweep. It shows the applied voltage and the measured current. When the current reaches compliance (here 90 µA) the voltage is automatically regulated down, which explains the part of the *I-V* curves where the current is rising while the voltage is falling. b) Voltage sweeps of four reference devices each with no RS medium (neither $PtSe_2$ nor polymer residue) on the three samples ($PtSe_2$-A, -B, and polymer). The Ohmic behavior (average resistances of four devices for each sample are included in the figure) and low resistances from -3 V to 3 V confirm that the observed RS is due to the active material and not due to the electrodes or any parasitic oxides.

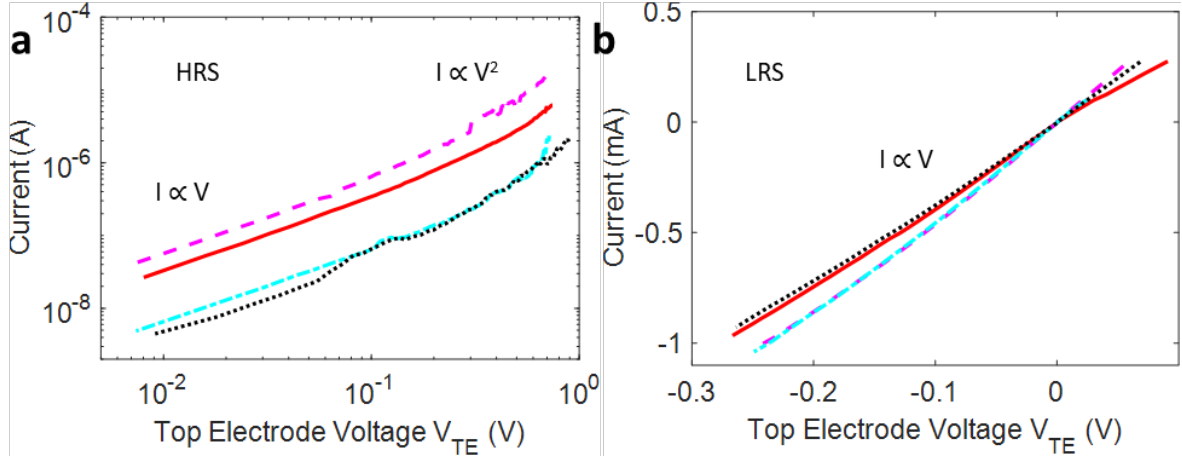

Figure S6: Additional plots of HRS and LRS for the modeling of the polymer sample. a) Median *I-V* characteristic of four additional devices in the HRS showing a linear and a quadratic slope. b) *I-V* characteristic of four additional devices in the LRS, showing linear behavior.

The experimental *I-V* characteristic (Fig. 3d) in the HRS is fitted using the phenomenological model equation described in the main text (equation 1). The parameters *µ*, *ε*, and *L* are unknown and can vary from cycle to cycle and with voltage polarity. Therefore, separate fitting is used for the positive and the negative voltage polarity. The resultant equations are $I = 2.7 \cdot 10^{-6}V + 3 \cdot 10^{-6}V^2$, and $I = 3 \cdot 10^{-6}V + 5 \cdot 10^{-6}V^2$ for forward and negative voltage polarity, respectively. To ensure consistency with the model the ratio of the fitting pre-factors is proportional to $\frac{32\pi^2 k_B T}{9q}$ following equation (1). The LRS is fitted using the Ohmic law $I = V/R$, where *R* is extracted to be 290 Ω.